\documentclass[a4paper,12pt]{article}

\usepackage[utf8]{inputenc}   
\usepackage[T1]{fontenc}       

\usepackage{graphicx}
\usepackage{geometry}
\usepackage{verbatim}
\usepackage{amsfonts,amssymb}
\usepackage[normalem]{ulem}
\usepackage{color}
\usepackage{varioref,float,amsmath,amsthm}
\usepackage{mathtools}
\usepackage{multirow}
\usepackage{algorithm}
\usepackage[version=4]{mhchem}
\usepackage{hyperref}
\usepackage{algorithmic}

\begin{document}

\begin{center}
.\\ \vspace{7cm}
\noindent\textbf{\LARGE
Fractional reversible chemical kinetics: Time-Driven Monte Carlo method and application to enzyme conformational dynamics
}\

\vspace{2cm}
{\small\noindent
{\bf Nelson H. T. Lemes} and {\bf Higor V. M. Ferreira}
} \
\vspace{2cm}
\end{center}

\noindent
\begin{center}
{\small
Laboratory of Mathematical-Chemistry, Institute of Chemistry, Universidade Federal de Alfenas (UNIFAL), Alfenas, MG, Brazil  \
}
\end{center}
\vspace{\stretch{1}}
\noindent{$^*$nelson.lemes@unifal-mg.edu.br}

\newpage
\subsubsection*{Abstract}
\small{
In this work, we present a fractional framework for reversible two-state chemical kinetics derived from the Continuous-Time Random Walk (CTRW) formalism with Mittag-Leffler waiting-time distributions. This derivation provides a microscopic basis for the fractional kinetic equations governed by the Caputo derivative. An exact analytical solution is obtained, yielding closed-form expressions for the time evolution of the state populations. To the best of our knowledge, a time-driven Monte Carlo algorithm is developed for the first time to simulate the irreversible underlying non-Markovian stochastic dynamics governed by a system of Caputo fractional differential equations. The proposed framework is applied to the conformational dynamics of the TbQSOX enzyme, whose relaxation exhibits power-law behavior. The fractional model accurately reproduces the experimental kinetics, providing an estimate of the fractional order $\alpha = 0.22$, which is consistent with the observed anomalous dynamics. \\
}
\newline
\noindent
{\bf Keywords:}
Fractional chemical kinetics;
Continuous-time random walk;
Caputo fractional derivative;
Time-driven Monte Carlo simulation;
Enzyme conformational dynamics.

\clearpage
\section{Introduction}

In 2018, Grossman-Haham and co-workers~\cite{grossman2018slow} investigated the conformational dynamics of the multidomain enzyme TbQSOX (Quiescin Sulfhydryl Oxidase). TbQSOX is an electron-transfer enzyme from the parasite \textit{Trypanosoma brucei} that primarily catalyzes the formation of disulfide bonds in proteins. In cells, this enzyme is localized in the Golgi apparatus and is also secreted into the extracellular environment. Multidomain enzymes are proteins composed of two or more functionally distinct domains within a single molecular entity. For TbQSOX, its function depends on transitions between open and closed conformations, enabling communication between the active sites and electron transfer across domains. These conformational transitions constitute the central focus of the aforementioned study.

In {chemical} kinetics, the investigation of reaction rates and molecular mechanisms is pivotal to elucidating how substrate concentration, temperature, and catalytic agents govern the velocity of chemical transformations.
The connection between reaction rates and molecular mechanisms is established by systems of Ordinary Differential Equations (ODEs).
Advancements in this field are essential for optimizing industrial processes across biotechnology, pharmaceuticals, and food sciences.

The ODE formalism is based on the fundamental principles of collision theory~\cite{Levine1987}.
Traditionally, transitions between two conformational states are described by systems of ODEs, which predict exponential relaxation from one state to the other until equilibrium is reached. In most practical applications, these equations are formulated using classical integer-order derivatives because of their straightforward interpretation as rates of change. However, such local operators have a limited ability to represent temperature gradients, diffusion, transport phenomena, and other nonlocal effects inherent to realistic reaction environments. Experimental evidence reported by Grossman-Haham \textit{et al.}~\cite{grossman2018slow} demonstrates that the transition between the two principal conformational states of TbQSOX, namely the closed state ($A$) and the open state ($B$), does not follow the conventional exponential relaxation. 

Using single-molecule Förster Resonance Energy Transfer (smFRET), the authors of Ref.~\cite{grossman2018slow} measured intermolecular distances and monitored structural changes in individual biomolecules as they freely diffused through a small observation volume, where the excitation laser was focused to excite fluorescent probes and detection efficiency was maximized. Combined with the Recurrence Analysis of Single Particles (RASP) technique, this approach enabled the monitoring of relaxation dynamics and interconversion kinetics by initially selecting only molecules in the closed state over timescales ranging from microseconds to hundreds of milliseconds. Experimental details can be found in Ref.~\cite{grossman2018slow}.

The authors explicitly identify TbQSOX as a ``remarkable exception'' to the classical rate-equation framework.
Specifically, Grossman-Haham and co-workers~\cite{grossman2018slow} demonstrated that the decay of the closed-state population follows a power law of the form
$p_{A}(t)\propto t^{-|\beta|}$. This finding challenges the long-standing paradigm of exponential kinetics~\cite{Levine1987}.

Despite the anomalous relaxation dynamics, the system eventually reaches a macroscopic equilibrium by continuously interconverting between the open and closed conformational states, although the pathway toward equilibrium is considerably more complex than predicted by classical kinetic models. In the absence of a substrate, approximately $88\%$ of the molecules reside in the open state, whereas only $12\%$ adopt the closed conformation~\cite{grossman2018slow}.

The article by Grossman-Haham and co-workers~\cite{grossman2018slow} adopts two distinct approaches to explain the experimental observations, neither of which employs the Caputo-derivative framework proposed in the present work.

The first is a purely empirical approach, in which the authors show that the decay of the closed-state population appears as a straight line on a logarithmic plot, following the relationship
$p_{A}(t)\propto t^{-|\beta|}$.
In addition to the linear regression on a $\log \times \log$ plot, they also evaluated alternative fitting functions, including the Kohlrausch--Williams--Watts (KWW) function~\cite{Ma2005} and a double-exponential model~\cite{Zwanzig1988,Berezhkovskii2011}. They concluded that the power-law model provided the statistically superior fit.

As a second approach, the authors sought to identify the physical origin of the observed behavior. They proposed that the domain motion should be described as a subdiffusive process evolving over a rugged free-energy landscape. Accordingly, classical rate equations were replaced by a diffusion equation governing the time evolution of the probabilities associated with the open and closed conformational states. Instead of assuming a constant diffusion coefficient, the model employs a time-dependent diffusion coefficient,
$D(t)\propto t^{\alpha-1}$,
with $\alpha=0.23$~\cite{Satija2017}.
Furthermore, the diffusion model requires an effective potential describing the enzyme's free-energy landscape along the distance coordinate used to monitor its structural dynamics. To estimate this potential, the authors employed experimentally measured distance distributions. The resulting diffusion equation was then solved numerically to obtain the probability density $p(r,t)$. The theoretical fraction of molecules in the closed state, $p_{A}(t)$, at each time $t$ was subsequently computed by integrating the probability density $p(r,t)$ over the corresponding region of configuration space. Finally, the model parameters were adjusted to reproduce the experimentally observed temporal decay curve.

The diffusion model used by Grossman-Haham and co-workers~\cite{grossman2018slow} is frequently presented in the literature as a governing equation for fractional Brownian motion (fBm).
However, the Continuous-Time Random Walk (CTRW) framework~\cite{montroll1965,scher1975,Metzler2000}, which leads to fractional diffusion equations, and fractional Brownian motion are based on fundamentally different physical assumptions, although both give rise to anomalous diffusion~\cite{Metzler2000,Sokolov2004}. 
This distinction has been extensively discussed in the anomalous diffusion literature and is one of the main reasons why CTRW and fractional Brownian motion are now recognized as belonging to distinct classes of non-Markovian stochastic processes.

The diffusion equation employed in Ref.~\cite{grossman2018slow} is a one-dimensional variant of the Fokker--Planck (or Smoluchowski) equation with a time-dependent diffusion coefficient. It is a linear, parabolic, non-autonomous partial differential equation with variable coefficients. Notably, both the spatial and temporal differential operators are identical to those appearing in the classical Smoluchowski equation. Consequently, this formulation represents a fundamentally different perspective from the one proposed in the present work. 

This distinction is particularly relevant because the CTRW derivation presented here provides an explicit physical basis for the fractional operators used in modeling chemical kinetics, whereas the approach based on a time-dependent diffusion coefficient, $D(t)$, provides an effective description of the dynamics without explicitly identifying the microscopic mechanism responsible for the observed anomalous behavior. To fit the experimental population decay data, the authors adopted the empirical relation $D(t)=D(0)(1+bt)^{-c}$,
where $D(0)$, $b$, and $c$ are empirical fitting parameters that control the temporal decay of the molecules.

The authors of Ref.~\cite{grossman2018slow} explicitly acknowledge both the theoretical and experimental limitations that restrict the scope of their conclusions. Although the power-law fit was found to be statistically superior, they recognize that this description remains purely empirical. Regarding the proposed physical model, they also point out the difficulty of determining the effective potential $V(r)$, acknowledging that the current understanding of the structural origin of dynamic disorder is still incomplete and that the proposed physical model has clear limitations.

Reference~\cite{Jangid2024} presents a recent theoretical study that seeks to explain why certain molecular systems do not exhibit classical exponential kinetics, following essentially the same conceptual framework adopted in Ref.~\cite{grossman2018slow}. In the work of Jangid and Chaudhury~\cite{Jangid2024}, the main innovation lies in deriving a diffusion coefficient that effectively embeds the complexity of the free-energy barrier into the diffusion term itself, in contrast to the approach of Grossman-Haham and co-workers, who employ an empirical functional form for $D(t)$ fitted directly to the experimental data.

In the present work, we propose replacing the diffusion-equation-based model employed in Ref.~\cite{grossman2018slow} to describe the anomalous decay of the closed state of the TbQSOX enzyme with a kinetic equation governed by non-integer-order Caputo derivatives.
Within our framework, anomalous effects are incorporated directly into the structure of the kinetic model through a fractional differential equation, rather than being introduced indirectly by means of a time-dependent diffusion coefficient in an otherwise classical diffusion equation.

Our approach provides a rigorous theoretical framework for the emergence of power-law kinetics in a two-state system.
Similarly, Ref.~\cite{agilan2025analysis} proposed a variable-order Caputo fractional enzyme-kinetics model incorporating constant time delays to more accurately capture memory effects and nonlocal behavior. Other recent studies that employ Caputo fractional-order models to investigate enzymatic reaction dynamics include Refs.~\cite{high2024performance,radhakrishnan2025systematic}.

Therefore, our work presents three key contributions. The first comprises the development of a stochastic algorithm based on a time-driven Monte Carlo method capable of directly reproducing the fractional dynamics described by the Caputo derivative. The second contribution is the use of the analytical solution for the fractional reversible kinetic model, which provides a rigorous benchmark for validating the proposed Monte Carlo algorithm.
Finally, the primary contribution of this work lies in the quantitative application of this formalism to the conformational dynamics of the TbQSOX enzyme. 

This study provides a unified framework integrating stochastic modeling, analytical derivation, computational simulation, and experimental validation, thereby offering a robust mechanistic interpretation of the anomalous kinetics observed in complex biomolecular systems, accurately capturing both the temporal evolution and the asymptotic equilibrium behavior. The present work constitutes a direct extension and an important application of the previous paper by Lemes et al.~\cite{TCAM}.

The remainder of this paper is organized as follows. Section~2 introduces the classical reversible two-state kinetic model and its fractional generalization derived from the Continuous-Time Random Walk (CTRW) framework, establishing the microscopic basis of the Caputo fractional operator. Section~3 presents an exact analytical solution of the fractional kinetic equations, providing closed-form expressions for the state populations and a theoretical benchmark for subsequent analyses. Section~4 describes the proposed time-driven Monte Carlo algorithm for simulating the underlying non-Markovian stochastic dynamics. Section~5 applies the proposed framework to the conformational dynamics of the TbQSOX enzyme, where numerical simulations and analytical predictions are compared with experimental observations. Finally, Section~6 summarizes the main findings and discusses the broader implications of the proposed framework for modeling reversible chemical kinetics with memory effects in complex biomolecular systems.

\section{Theoretical Fundamentals}

To establish a rigorous physical basis for the numerical methodology developed in this work, we first revisit the connection between the microscopic CTRW framework and the corresponding macroscopic fractional kinetic equations. Although this connection is well known in fractional dynamics, its explicit derivation is essential in the present context because it demonstrates that the fractional operators employed throughout this study arise naturally from microscopic trapping phenomena characterized by heavy-tailed waiting-time statistics, rather than from phenomenological curve-fitting procedures.

\subsection{Derivation of fractional reversible kinetics from the CTRW formalism}

Throughout this derivation, we consider a decoupled continuous-time random walk (CTRW), namely a stochastic process in which the waiting times and transition events are statistically independent. A two-state reversible process is described by the chemical equation:

$$A \rightleftharpoons B,$$

\noindent
where the molecule---for example, the enzyme TbQSOX~\cite{grossman2018slow}---may occupy either conformational state $A$ (closed) or conformational state $B$ (open) and undergo transitions between them.
Let $P_A(t)$ and $P_B(t)$ denote the probabilities of finding the system in states $A$ and $B$, respectively, at time $t \ge 0$. Probability conservation requires
\begin{equation}
P_A(t) + P_B(t) = 1,
\end{equation}
for all times.

The CTRW description requires the specification of waiting-time probability density functions (PDFs), $\psi_A(t)$ and $\psi_B(t)$, where $\psi_A(t)$ denotes the probability density for leaving state $A$ after a residence time $t$, and $\psi_B(t)$ is defined analogously for state $B$.
Associated with these densities are the survival probabilities.
The survival function, $\Psi_A(t)$, defined as the probability that the transition has not yet occurred up to time $t$, is given by the expression:
\begin{equation}
\Psi_A(t) = \int_t^\infty \psi_A(t'),dt' = 1 - \int_0^t \psi_A(t'),dt'.
\label{PsiAt}
\end{equation}
The survival function $\Psi_B(t)$ is defined analogously for state $B$:
\begin{equation}
\Psi_B(t) = 1 - \int_0^t \psi_B(t'),dt'.
\end{equation}
Thus, $\Psi_A(t)$ or $\Psi_B(t)$ represent the probabilities that the system remains in states $A$ and $B$, respectively, without undergoing a transition up to time $t$.

In this problem, the probability $P_A(t)$ decreases due to $A \rightarrow B$ transitions and increases due to reverse $B \rightarrow A$ transitions.
In the CTRW framework, the fluxes between states A and B depend on the prior history of the system. Defining the generalized fluxes as
\begin{equation}
J_{A\rightarrow B}(t) = \int_0^t K_A(t-t') P_A(t'),dt',
\end{equation}
and
\begin{equation}
J_{B\rightarrow A}(t) = \int_0^t K_B(t-t') P_B(t'),dt',
\end{equation}
probability conservation for state $A$ yields
\begin{equation}
\frac{dP_A(t)}{dt} = -J_{A\rightarrow B}(t) + J_{B\rightarrow A}(t).
\label{eq7}
\end{equation}

The expression
\begin{equation}
J_{A\rightarrow B}(t) = \int_0^t K_A(t-t') P_A(t'),dt'
\label{generalized_flux}
\end{equation}
is a non-Markovian generalization of the conventional kinetic law. In the Markovian case, the transition flux is proportional to the instantaneous population in state $A$,
\begin{equation}
J_{A\rightarrow B}(t) = k_{AB} P_A(t).
\end{equation}
In CTRW, however, the probability of a transition occurring at time $t$ depends not only on the population present in $A$ at that instant, but also on the time systems have spent in state $A$. Therefore, the flux must incorporate the temporal history of $P_A$.
This temporal dependence is described by a convolution with the memory kernel $K_A(t)$, as in Eq.~(\ref{generalized_flux}). The factor $P_A(t')$ represents the probability of the system being in state $A$ at an earlier time $t'$, whereas $K_A(t-t')$ weights the contribution of this population to a transition occurring after the elapsed time interval $t-t'$.

Substituting the expressions for the fluxes into Equation (\ref{eq7}) yields
\begin{equation}
\frac{dP_A(t)}{dt} = -\int_0^t K_A(t-t')P_A(t'),dt' + \int_0^t K_B(t-t')P_B(t'),dt'.
\label{rev_master_A_derivation}
\end{equation}
Applying the Laplace transform to Eq.~(\ref{rev_master_A_derivation}) yields
\begin{equation}
s\hat{P}_A(s) - P_A(0) = -\hat{K}_A(s)\hat{P}_A(s) + \hat{K}_B(s)\hat{P}_B(s).
\label{rev_laplace_A}
\end{equation}
The first term on the right-hand side of Eq.~(\ref{rev_laplace_A}) represents the loss of probability associated with the transition $A \rightarrow B$, whereas the second term represents the probability gain in state $A$ caused by the reverse transition $B \rightarrow A$. In an irreversible process, this last term is absent because species $A$ is not replenished~\cite{TCAM}.
By symmetry, the corresponding balance equation for state $B$ is
\begin{equation}
s\hat{P}_B(s) - P_B(0) = \hat{K}_A(s)\hat{P}_A(s) - \hat{K}_B(s)\hat{P}_B(s).
\label{rev_laplace_B}
\end{equation}
These equations constitute the Generalized Master Equations (GMEs) governing the reversible CTRW process.

Adding Eqs.~(\ref{rev_laplace_A}) and (\ref{rev_laplace_B}) gives
\begin{equation}
s\left[\hat{P}_A(s) + \hat{P}_B(s)\right] = P_A(0) + P_B(0) = 1,
\end{equation}
which implies probability conservation $P_A(t) + P_B(t) = 1$.

We conveniently introduce the memory kernel in the Laplace domain, associated with each transition direction, through the following definitions:
\begin{subequations}
\begin{align}
\hat{K}_A(s) &:= \frac{s\hat{\psi}_A(s)}{1-\hat{\psi}_A(s)}, \\
\hat{K}_B(s) &:= \frac{s\hat{\psi}_B(s)}{1-\hat{\psi}_B(s)}.
\end{align}
\end{subequations}
Here, $\hat{K}_A(s)$ describes the memory associated with the transition $A \rightarrow B$, whereas $\hat{K}_B(s)$ describes that associated with $B \rightarrow A$.
In the above equations, $\hat{\psi}_A(s)$ is the Laplace transform of the waiting-time probability density function $\psi_A(t)$, and
$\hat{\psi}_B(s)$ is defined analogously for state $B$.

The convenience of defining the memory kernel in Laplace space stems from the exact equivalence between the microscopic stochastic Continuous-Time Random Walk (CTRW) theory and the differential Generalized Master Equation (GME). This relationship ensures that the choice of $\hat{K}_A(s)$ exactly reproduces the waiting-time statistics $\hat{\psi}_A(s)$ of the microscopic model. The mathematical origin of this expression was initially demonstrated for the irreversible process $A \rightarrow B$~\cite{TCAM}.

Consider now the non-Markovian case, in which the waiting-time density exhibits a heavy tail. In this context, one assumes an asymptotic form compatible with power-law distributions, characterized in the Laplace domain by
\begin{equation}
\hat{\psi}_A(s) = \frac{1}{1 + (\tau_A s)^\alpha},
\qquad
0 < \alpha < 1.
\label{psi1}
\end{equation}
The choice of $\hat{\psi}_A(s)$ corresponds in the time domain to waiting-time statistics governed by a two-parameter Mittag-Leffler function~\cite{gorenflo2014}:
\begin{equation}
\psi_A(t) = \frac{t^{\alpha-1}}{\tau_A^\alpha} E_{\alpha,\alpha}\left(-\left(\frac{t}{\tau_A}\right)^\alpha\right).
\end{equation}
The $E_{\alpha, \beta}(z)$ function is defined by

\begin{equation}
E_{\alpha, \beta}(z) = \sum_{k=0}^{\infty} \frac{z^k}{\Gamma(\alpha k + \beta)},
\end{equation}

\noindent
where $z \in \mathbb{C}$ is a complex variable, $\alpha$ and $\beta$ are complex parameters constrained by $\mathrm{Re}(\alpha) > 0$ for convergence, and $\Gamma(\cdot)$ denotes the classical Gamma function.
In this case, the survival function associated with the waiting-time density $\psi_A(t)$ is
\begin{equation}
\Psi_A(t) = E_{\alpha,1}\left(-\left(\frac{t}{\tau_A}\right)^\alpha\right),
\end{equation}
with the asymptotic limit for $t \to \infty$ given by
\begin{equation}
\Psi_A(t) \sim \frac{\tau_A^\alpha}{\Gamma(1-\alpha)} t^{-\alpha}.
\end{equation}
The assumption that both states are characterized by the same fractional exponent $\alpha$ is generally necessary to ensure a well-defined asymptotic population ratio.

Inserting $\hat{\psi}_A(s)$ and $\hat{\psi}_B(s)$ into the definition of the memory kernels leads to
\begin{subequations}\label{eq:sistemax}
\begin{align}
\hat{K}_A(s) &= \tau_A^{-\alpha} s^{1-\alpha}, \label{eq:sistemax_a} \\
\hat{K}_B(s) &= \tau_B^{-\alpha} s^{1-\alpha}. \label{eq:sistemax_b}
\end{align}
\end{subequations}
Using these kernels in the generalized master equations, one obtains
\begin{subequations}\label{eq:sistemay}
\begin{align}
s^\alpha \hat{P}_A(s) - s^{\alpha-1} P_A(0) &= -\tau_A^{-\alpha} \hat{P}_A(s) + \tau_B^{-\alpha} \hat{P}_B(s), \label{eq:sistemay_a} \\
s^\alpha \hat{P}_B(s) - s^{\alpha-1} P_B(0) &= \tau_A^{-\alpha} \hat{P}_A(s) - \tau_B^{-\alpha} \hat{P}_B(s). \label{eq:sistemay_b}
\end{align}
\end{subequations}

In the present work, the temporal memory of the system is modeled through the Caputo fractional derivative.
In the fractional model, ${}_0D_t^\alpha\phi(t)$ denotes the Caputo fractional derivative of order $\alpha$. For $0<\alpha<1$, let $\phi=f:\mathbb{R}\to\mathbb{R}$ be locally absolutely continuous on $(-\infty,t]$ and suppose that $f'$ satisfies the integrability condition $\int_{-\infty}^{t}\vert{}f'(\tau)\vert{}(t-\tau)^{-\alpha}\, d\tau<\infty$. A Caputo-type fractional derivative with memory extending over the entire past is then defined by~\cite{podlubny1999,kilbas2006,diethelm2010}:

\begin{equation}
{}_{-\infty}D_{t}^{\alpha}f(t) := \frac{1}{\Gamma(1-\alpha)} \int_{-\infty}^{t} \frac{f'(\tau)}{(t-\tau)^{\alpha}} , d\tau, \qquad 0<\alpha<1.
\end{equation}
This definition is free from an arbitrarily prescribed finite lower terminal and, consequently, incorporates the complete temporal history of the process. In particular, it is compatible with the translation-invariant formulation of fractional differentiation and avoids the dependence on an externally selected initial time. If, in addition, $f$ has its support contained in $[0,\infty)$---namely, $f(\tau)=0$ for $\tau<0$---with $f$ being locally absolutely continuous across $t=0$ (so that no impulsive contribution occurs at the origin), then, for every $t>0$:

\begin{equation}
{}_{-\infty}D_{t}^{\alpha}f(t) = {}_{0^+}D_{t}^{\alpha}f(t),
\end{equation}
which is precisely the classical Caputo derivative of order $\alpha$ with a lower terminal of $0^+$. In this scenario, the function $f(\tau)$ is nonzero only for $\tau \geq 0$ and vanishes identically outside this domain. Hence, the standard Caputo operator is recovered as the causal restriction of the full-history operator whenever the prehistory of $f$ vanishes before $t=0$.
This definition clearly highlights the nonlocal character of the operator, since the fractional rate of change at $t$ depends on the entire previous history of the function $f(\tau)$ over the interval $[0,t]$, weighted by a power-law memory kernel.

The Caputo formulation is particularly suitable for physical applications because it allows the use of classical initial conditions expressed in terms of integer-order derivatives~\cite{podlubny1999}. In the context of stochastic chemical kinetics, this operator introduces a history-dependent contribution to the reaction dynamics, effectively accounting for memory effects that arise from non-exponential waiting-time statistics.
If $f(t)$ is a sufficiently regular function for $t \ge 0$, then the Laplace transform of the Caputo derivative is given by
\begin{equation}
\mathcal{L}\{{}_{0^+}D_t^\alpha f(t)\}(s) = s^\alpha \hat{f}(s) - s^{\alpha-1}f(0),
\end{equation}
for $0 < \alpha \le 1$, where $\hat{f}(s) = \mathcal{L}\{f(t)\}(s)$~\cite{caputo1967,podlubny1999,kilbas2006,diethelm2010}.

Furthermore, we assume that the state probabilities $P_A(t)$ and $P_B(t)$ are absolutely continuous on every finite interval $[0,T]$ with $T > 0$, ensuring the existence of the Caputo fractional derivative of order $0 < \alpha < 1$.
Returning to the time domain from Eqs.~(\ref{eq:sistemay}), one finally obtains the fractional kinetic equations:
\begin{subequations}\label{eq:sistemaz}
\begin{align}
{}_{0^+}D_t^\alpha P_A(t) &= -\tau_A^{-\alpha} P_A(t) + \tau_B^{-\alpha} P_B(t), \label{eq:sistemaz_a} \\
{}_{0^+}D_t^\alpha P_B(t) &= \tau_A^{-\alpha} P_A(t) - \tau_B^{-\alpha} P_B(t). \label{eq:sistemaz_b}
\end{align}
\end{subequations}
Introducing the apparent fractional rate constants $k_1^\alpha = \tau_A^{-\alpha}$ and $k_2^\alpha = \tau_B^{-\alpha}$, the system assumes the compact form
\begin{subequations}\label{eq:sistemaw}
\begin{align}
{}_{0^+}D_t^\alpha P_A(t) &= -k_1^\alpha P_A(t) + k_2^\alpha P_B(t), \label{eq:sistemaw_a} \\
{}_{0^+}D_t^\alpha P_B(t) &= k_1^\alpha P_A(t) - k_2^\alpha P_B(t). \label{eq:sistemaw_b}
\end{align}
\end{subequations}
The classical Markovian limit is recovered when $\alpha = 1$, corresponding to exponential waiting-time statistics and memoryless reaction dynamics.

In this case, dimensional equality between both sides of the equation is ensured, with the reaction rate constant having the dimension $[k] =  \mathsf{T}^{-1}$. On the left-hand side, we have $\left[ d^\alpha [A]/dt^\alpha \right] = \frac{[[A]]}{[t]^\alpha} = \mathsf{C} \cdot \mathsf{T}^{-\alpha}$ and, on the right-hand side, we have
$[k]^\alpha \cdot [[A]]= \left( \mathsf{T}^{-1} \right)^\alpha \cdot \mathsf{C} = \mathsf{C} \cdot \mathsf{T}^{-\alpha}$,  
where 
$\mathsf{T}$ and $\mathsf{C}$ represent the physical dimensions of time and concentration, respectively.
By raising the classical rate constant $k$ to the power of $\alpha$, the term $k^\alpha$ acquires the dimension $[k^\alpha] = \mathsf{T}^{-\alpha}$. This exactly compensates for the power of time introduced by the fractional operator, ensuring the dimensional consistency of the equation without altering the original dimension of the rate constant $k$.

The derivation thus demonstrates that reversible fractional kinetics emerges directly from microscopic stochastic assumptions. The Caputo derivative is not introduced phenomenologically; instead, it arises naturally from the heavy-tailed Mittag-Leffler residence-time statistics governing both forward and reverse transitions.

\subsection{Exact solution of the two-state model}

Obtaining analytical solutions for kinetic systems is not always feasible. In this context, one of the contributions of this work is the derivation of the exact analytical solution for a two-state reversible kinetic model governed by the Caputo fractional derivative. Beyond merely providing an explicit expression for the temporal evolution of the molecular fractions, this analytical solution plays a pivotal role in the development of this study, serving as a benchmark for validating the proposed stochastic algorithm and as a direct model for estimating kinetic parameters from experimental data.

Consider a reversible process $A \rightleftharpoons B$. In the thermodynamic limit, the occupation probability of a given state coincides with the fraction of the population occupying that state.
This correspondence is widely employed in chemical kinetics, providing the basis for transitioning from the microscopic description, expressed in terms of the probabilities $P_A(t)$ and $P_B(t)$, to the macroscopic description formulated in terms of the molecular fractions $f_A(t)$ and $f_B(t)$.

In the present work, the proportionality relation $f_A(t) \propto P_A(t)$ is further supported by the design of the experimental data reported by Grossman-Haham \textit{et al.}~\cite{grossman2018slow}. Specifically, the conformational relaxation data shown in Figure~2A were obtained by statistically aligning single-molecule trajectories acquired using smFRET, where the time origin ($t=0$) is rigorously defined as the exact instant at which each molecule enters the conformational state under investigation.

The equations in \eqref{eq:sistemaw} are coupled because the time evolution of each species depends simultaneously on the molecular fractions of both components, thereby preventing each differential equation from being solved independently. A standard approach to decouple the system is to incorporate the stoichiometric constraints of the reaction. Assuming $f_B(0) = 0$, the molecular fractions can be expressed in terms of a single reaction-progress variable $x(t)$ as
\begin{subequations}\label{eq:sistema1w}
\begin{align}
f_A(t) &= f_A(0) - x(t), \label{eq:sistema1w_a} \\
 f_B(t) &= x(t). \label{eq:sistema1w_b}
\end{align}
\end{subequations}

Let $c \in \mathbb{R}$ be a constant function defined on $t \in [0, \infty)$. By the linearity of the Caputo fractional derivative,
\begin{equation}
{}_{0^+}D_t^\alpha \bigl(c - x(t)\bigr)
= {}_{0^+}D_t^\alpha c
- {}_{0^+}D_t^\alpha x(t).
\end{equation}
Since the Caputo operator ${}_{0^+}D_t^\alpha$ acts strictly on the interval $\tau \in (0, t)$ where $\frac{d}{d\tau}c(t) = 0$, the constant term vanishes (${}_{0^+}D_t^\alpha c(t) = 0$ for $0 < \alpha < 1$), leading to
\begin{equation}
D_t^\alpha \bigl(c - x(t)\bigr)
=
-D_t^\alpha x(t).
\end{equation}

Applying this property to the differential equation (\ref{eq:sistemaw_a}) governing species $A$ yields
\begin{align}
-D_t^\alpha x(t)
&=
-k_1^\alpha \bigl(f_A(0)-x(t)\bigr)
+k_2^\alpha x(t),
\\
&=
-k_1^\alpha f_A(0)
+k_1^\alpha x(t)
+k_2^\alpha x(t),
\\
&=
-k_1^\alpha f_A(0)
+\left(k_1^\alpha+k_2^\alpha\right)x(t),
\\
D_t^\alpha x(t)
&=
k_1^\alpha f_A(0)
-\left(k_1^\alpha+k_2^\alpha\right)x(t).
\end{align}
The resulting equation can be written in the compact form
\begin{equation}
D_t^\alpha x(t)=a-bx(t),
\label{eq:11}
\end{equation}
where $a=k_1^\alpha f_A(0)$ and $b=k_1^\alpha+k_2^\alpha$. The original coupled system is thus reduced to a single linear fractional differential equation of order $\alpha$. 

Applying the Laplace transform $\mathcal{L}\{\cdot\}$ of the Eq. (\ref{eq:11}), under the initial condition $x(0)=0$, gives
\begin{equation}
s^\alpha X(s)=\frac{a}{s}-bX(s),
\end{equation}
from which
\begin{equation}
X(s)=\frac{a}{s(s^\alpha+b)}.
\end{equation}
Taking the inverse Laplace transform yields the solution in terms of the Mittag--Leffler function $E_\alpha$:
\begin{equation}
x(t)=\frac{a}{b}\left[1-E_\alpha(-bt^\alpha)\right].
\end{equation}
Finally, restoring the original physical variables, the reaction-progress function is given by
\begin{equation}
x(t)
=
\frac{k_1^\alpha f_A(0)}{k_1^\alpha+k_2^\alpha}
\left[
1-
E_\alpha!\left(-(k_1^\alpha+k_2^\alpha)t^\alpha\right)
\right].
\label{eqx}
\end{equation}

From $x(t)$, the explicit time-dependent molecular fraction profiles of species $A$ and $B$, Eqs. (\ref{eq:sistema1w_a}) and
(\ref{eq:sistema1w_b}), are readily obtained:
\begin{align}
f_A(t)
&=
f_A(0)-x(t)
= f_A(0)
- \frac{k_1^\alpha f_A(0)\left[1-E_\alpha\bigl(-(k_1^\alpha+k_2^\alpha)t^\alpha\bigr)\right]}
{k_1^\alpha+k_2^\alpha}
\nonumber\\
&=
\frac{f_A(0)(k_1^\alpha+k_2^\alpha)
k_1^\alpha f_A(0)
\left[
1-E_\alpha\bigl(-(k_1^\alpha+k_2^\alpha)t^\alpha\bigr)
\right]}
{k_1^\alpha+k_2^\alpha}
\nonumber\\
&=
\frac{
f_A(0)k_1^\alpha
+f_A(0)k_2^\alpha
-k_1^\alpha f_A(0)
+k_1^\alpha f_A(0)
E_\alpha\bigl(-(k_1^\alpha+k_2^\alpha)t^\alpha\bigr)}
{k_1^\alpha+k_2^\alpha}
\nonumber\\
&=
\frac{
f_A(0)
\left[
k_2^\alpha
+
k_1^\alpha
E_\alpha\bigl(-(k_1^\alpha+k_2^\alpha)t^\alpha\bigr)
\right]}
{k_1^\alpha+k_2^\alpha},
\end{align}
whereas
\begin{equation}
f_B(t)
= x(t)
\frac{
k_1^\alpha f_A(0)
\left[
1-
E_\alpha\bigl(-(k_1^\alpha+k_2^\alpha)t^\alpha\bigr)
\right]}
{k_1^\alpha+k_2^\alpha}.
\end{equation}

Considering that the long-time asymptotic behavior of the Mittag--Leffler function is given by\cite{gorenflo2014}
\begin{equation}
E_{\alpha,1}(-b t^\alpha)
\sim
\frac{1}{b\Gamma(1-\alpha)}\,t^{-\alpha},
\end{equation}
which implies that
\begin{equation}
\lim_{t\to\infty}E_{\alpha,1}(-b t^\alpha)=0,
\end{equation}
the system consequently approaches a well-defined asymptotic kinetic equilibrium, whose equilibrium constant satisfies
\begin{equation}
K
= \frac{f_B(\infty)}{f_A(\infty)}=
\frac{k_1^\alpha}{k_2^\alpha}=
\left(\frac{k_1}{k_2}\right)^\alpha.
\label{equilibrio}
\end{equation}
For the particular case $\alpha=1$, this expression reduces exactly to the classical first-order kinetic result,
$K=\frac{k_1}{k_2}$.

\subsection{Time-driven Monte Carlo simulation}

A main methodological contribution of this work is the development of a time-driven simulation algorithm grounded in the continuous-time random walk (CTRW) formalism, where probabilities are computed from Mittag-Leffler-type distributions. Unlike classical implementations based on Markovian processes, the proposed algorithm explicitly preserves the temporal memory inherent to the stochastic process, naturally reproducing the non-Markovian dynamics associated with the Caputo derivative.

Monte Carlo methods became established as a general computational methodology during the 1940s. Researchers such as Stanislaw Ulam and John von Neumann recognized that complex problems involving neutron transport, diffusive phenomena, and stochastic systems could be effectively addressed through systematic random sampling, particularly when conventional analytical approaches proved intractable. Subsequent developments by Nicholas Metropolis and Enrico Fermi consolidated these ideas into a coherent computational framework~\cite{metropolis1949,metropolis1987,hammersley1964,ehrenfest1907,kac1947}.

In the context of chemical kinetics, a major theoretical advance was introduced by McQuarrie and co-workers, who formulated the stochastic dynamics of chemical reactions in terms of Poisson processes~\cite{mcquarrie1967,vanKampen2007}. Their work also explored temporal discretization as a practical computational approximation, thereby establishing the conceptual foundations of what is now known as the \emph{time-driven Monte Carlo} method. Within this framework, time is partitioned into uniform intervals ($\Delta t$), and the system evolves through probabilistic updates performed at each time step. Each simulated entity---whether a molecule, particle, radioactive nucleus, or reactive site---is assigned a transition probability ($p$), which is directly related to a kinetic rate constant ($\lambda$). A uniformly distributed random number ($r\in(0,1)$) is then generated for each entity. A transition occurs if $r<p$; otherwise, the entity remains unchanged during the current time interval. After all entities have been evaluated, the global state of the system is updated, and the simulation clock is advanced by $\Delta t$.
This algorithmic approach is well established in the chemical kinetics literature and has been extensively explored from a pedagogical perspective, serving as an accessible framework for introducing stochastic reaction dynamics in academic contexts. However, the generalization of this time-driven Monte Carlo scheme to the fractional-order regime remains relatively scarce in the literature.

To the best of our knowledge, Reference~\cite{TCAM} presents the first systematic study of a time-driven Monte Carlo simulation that reproduces the non-Markovian dynamics of an irreversible process associated with the Caputo derivative. In this work, we generalize the results of that previous study to a reversible process and apply it to experimental data (Section~5), obtained from Reference~\cite{grossman2018slow}, with relative success.
It is important to highlight that, although the generalization to the reversible case is straightforward, it is not a trivial extension of the work in Reference~\cite{TCAM}. In this case, a different strategy had to be employed to arrive at the expression for the transition probability than that used in Section~2.3 of Reference~\cite{TCAM}.


To rigorously reproduce the macroscopic coupled dynamics while preserving the correct asymptotic kinetic equilibrium, $K=(k_1/k_2)^\alpha$, the time evolution of the individual states cannot be modeled as completely independent processes. Instead, the system must evolve according to a single global relaxation process. As shown in Equation~(\ref{eqx}), the effective relaxation rate of the coupled system is defined as
\begin{equation}
\lambda = k_1^\alpha + k_2^\alpha.
\end{equation}
Therefore, the corresponding global survival probability is given by
 \begin{equation}
\Psi_{\mathrm{sys}}(t)
=
E_{\alpha,1}\left(-\lambda t^\alpha\right).
\label{mmm}
\end{equation}

The associated waiting-time probability density function is obtained from the survival probability as
$\psi_{\mathrm{sys}}(t)=
-\frac{d\Psi_{\mathrm{sys}}(t)}{dt}$,
as previously established in Equation~(\ref{PsiAt}). The function $\psi_{\mathrm{sys}}(t)$ represents the unconditional probability density that a reaction event occurs exactly at time $t$. The negative sign arises because the survival probability is a monotonically decreasing function of time, whereas a probability density must remain non-negative.

Consider that the occurrence of event $A \rightarrow B$ in the interval $(t, t + \Delta t]$ is defined as $\mathcal{A}$ and that remaining in state $B$ up to time $t$ is defined as $\mathcal{B}$. Therefore, the probability of event $\mathcal{A}$ occurring, given that event $\mathcal{B}$ has already occurred, is defined by:
\begin{equation}
P(\mathcal{A} \mid \mathcal{B}) = \frac{P(\mathcal{A} \cap \mathcal{B})}{P(\mathcal{B})}
\end{equation}
If the reaction time $T$ lies within the interval $(t, t + \Delta t]$, then $\mathcal{A} \subseteq \mathcal{B}$ and, therefore, $\mathcal{A} \cap \mathcal{B} = \mathcal{A}$. Thus,
 \begin{equation}
H_{\mathrm{sys}}(t)=
P(t<T\le t+\Delta t \mid T>t)
= \frac{P(t<T\le t+\Delta t)}{P(T>t)}
= \frac{\psi_{\mathrm{sys}}(t)}
{\Psi_{\mathrm{sys}}(t)}=-
\frac{1}{\Psi_{\mathrm{sys}}(t)}
\frac{d\Psi_{\mathrm{sys}}(t)}{dt}.
\label{eqnome}
\end{equation}
 To evaluate the time derivative in Eq. (\ref{eqnome}) we employ the differentiation identity of the Mittag--Leffler function,
\begin{equation}
\frac{dE_{\alpha,1}(x)}{dx}
=
\frac{E_{\alpha,0}(x)}{\alpha x}.
\end{equation}
\noindent
Substituting $x=-\lambda t^\alpha$ and applying the chain rule yields
\begin{equation}
\frac{d}{dt}
E_{\alpha,1}\left(-\lambda t^\alpha\right)
= -\lambda t^{\alpha-1}
E_{\alpha,\alpha}\left(-\lambda t^\alpha\right),
\end{equation}
\noindent
from which the global effective hazard function becomes
\begin{equation}
H_{\mathrm{sys}}(t)
=
\frac{1}{t}
\left|
\frac{
E_{\alpha,0}\left(-\lambda t^\alpha\right)}
{
E_{\alpha,1}\left(-\lambda t^\alpha\right)}
\right|.
\label{complica}
\end{equation}

Our working hypothesis is that the effective probability functions governing transitions from states $A$ and $B$ are defined as proportional fractions of the global hazard, weighted by the corresponding fractional kinetic rates:
\begin{subequations}\label{eq:sistemag1}
\begin{align}
h_A(t)
&=
\frac{k_1^\alpha}{\lambda}
H_{\mathrm{sys}}(t), \label{eq:sistemag1_a} \\
h_B(t)
&=
\frac{k_2^\alpha}{\lambda}
H_{\mathrm{sys}}(t). \label{eq:sistemag1_b}
\end{align}
\end{subequations}
Therefore, for a sufficiently small time increment $\Delta t$, the first-order transition probabilities at each simulation step are approximated by:
\begin{subequations}\label{eq:sistemar1}
\begin{align}
P_{A\rightarrow B}(t)
&=
h_A(t) \Delta t, \label{eq:sistemar1_a} \\
P_{B\rightarrow A}(t)
&=
h_B(t) \Delta t. \label{eq:sistemar1_b}
\end{align}
\end{subequations}
The working hypothesis is confirmed by agreement with the analytical results.

Furthermore, this globally coupled formulation guarantees that the ratio
\begin{equation}
\frac{P_{A\rightarrow B}(t)}{P_{B\rightarrow A}(t)}=\frac{h_A(t)}{h_B(t)}
\left(\frac{k_1}{k_2}\right)^\alpha
\end{equation}
remains constant throughout the simulation, thereby preserving the asymptotic equilibrium predicted by the exact analytical solution, Equation~(\ref{equilibrio}).


\subsection{Alternative time-driven algorithm}

Although the time-driven algorithm introduced in the previous section provides an adequate stochastic representation of the fractional dynamics, its numerical implementation becomes increasingly demanding as the anomalous exponent $\alpha$ decreases. From Equations~(\ref{eq:sistemag1}) and~(\ref{eq:sistemar1}), the transition probabilities are proportional to the global effective hazard function:
\begin{equation}
P(t) \propto H_{\mathrm{sys}}(t) \Delta t.
\end{equation}
Using the asymptotic expansion of the Mittag--Leffler function for short times, $E_{\alpha,0} \sim -\lambda t^\alpha / \Gamma(\alpha)$ and $E_{\alpha,1} \sim 1$~\cite{podlubny1999,gorenflo2014,kilbas2006}, one obtains
\begin{equation}
H_{\mathrm{sys}}(t) \sim \lambda t^{\alpha-1} / \Gamma(\alpha) \to \infty,
\end{equation}
when $t \to 0$. For moderate subdiffusion (e.g., $\alpha = 0.8$), this weak integrable singularity does not compromise the numerical convergence of the algorithm. However, for strongly anomalous dynamics (e.g., $\alpha = 0.2$), the divergence near the origin becomes sufficiently pronounced that an extremely small time step is required.
Consequently, advancing the simulation over experimentally relevant times may require billions of time steps, whereas increasing $\Delta t$ leads to transition probabilities that no longer provide an accurate approximation of the underlying continuous-time process.

To overcome this numerical limitation while preserving the stochastic interpretation of the model over the entire interval $0 < \alpha \le 1$, it is convenient to reformulate the transition probability directly in terms of the survival function, thereby eliminating the explicit evaluation of the singular hazard function. The probability that the reaction occurs during the subsequent interval $(t, t + \Delta t]$, conditioned on survival until time $t$, is given by Eq. (\ref{eqnome}). 
Since
\begin{equation}
P(T > t) = \Psi_{\mathrm{sys}}(t),
\end{equation}
and
\begin{equation}
P(t < T \le t + \Delta t) = \Psi_{\mathrm{sys}}(t) - \Psi_{\mathrm{sys}}(t + \Delta t),
\end{equation}
the exact conditional transition probability becomes
\begin{equation}
H_{\mathrm{sys}}(t) 
= 1 - \frac{\Psi_{\mathrm{sys}}(t + \Delta t)}{\Psi_{\mathrm{sys}}(t)},
\label{eq:Pexact_global}
\end{equation}
where $\Psi_{\mathrm{sys}}(t)$ is given by Eq.~(\ref{mmm}).
Finally, the transition probabilities associated with each reaction channel are obtained by partitioning the global probability according to the same fractional kinetic weights introduced in Equation~(\ref{eq:sistemag1}).
Now $H_{\mathrm{sys}}(t)$ is given by
\begin{equation}
H_{\mathrm{sys}}(t) = 1 - \frac{E_{\alpha,1}\left(-\lambda (t+\Delta t)^\alpha\right)}{E_{\alpha,1}\left(-\lambda t^\alpha\right)},
\label{eq:Pexact_global2}
\end{equation}
instead of Eq.~(\ref{complica}). 
Therefore, the proposed reformulation eliminates the numerical instability associated with the short-time singularity while preserving the stochastic interpretation and the correct asymptotic equilibrium, yielding excellent agreement with the exact analytical solution over the entire range $0<\alpha\le1$.

\section{Results and discussion}

To validate the robustness and practical applicability of the developed fractional stochastic formalism, we apply the proposed model to fit and interpret the experimental data reported by Grossman-Haham et al.~\cite{grossman2018slow}. The quantitative application of this framework to an empirical system constitutes an additional contribution of this study.

Reference~\cite{grossman2018slow} provides the temporal evolution of the population fraction of closed conformers (A). The experimental data reveal an anomalous kinetic profile that sharply deviates from the conventional exponential description characteristic of classical kinetics. Furthermore, the study demonstrates that the open state (B) is the predominant conformational species at equilibrium, accounting for approximately 88\% of the total population. 
This asymptotic thermodynamic distribution imposes strict constraints on our model parameters, fixing the ratio of the rate constants such that the long-time limit ($t \to \infty$) faithfully reproduces the experimentally observed population fractions. The experimental data were digitized from the figures reported by Grossman-Haham et al.~\cite{grossman2018slow} using \textit{WebPlotDigitizer} (Version 4.7), developed by Rohatgi~\cite{Rohatgi2024}. The numerical evaluation of the Mittag--Leffler function was performed using the MATLAB routine \texttt{mlf} developed by Podlubny and Kacenak~\cite{PodlubnyMLF}.

The parameter optimization procedure was formulated as a nonlinear least-squares problem to determine the parameter vector $\mathbf{p}=(f_A(0),k_1,k_2,\alpha)$ that best reproduces the experimental molecular fraction $f_A^{\mathrm{exp}}(t_i)$ using the fractional kinetic model $f_A^{\mathrm{model}}(t_i;\mathbf{p})$. The implementation utilized the \texttt{lsqnonlin} solver in MATLAB~\cite{OptimizationToolbox2024, Coleman1996,Coleman1994,Levenberg1944,Marquardt1963,More1977}, which minimizes the residuals between the experimental data and model predictions, augmented by a penalty term designed to approximately enforce the thermodynamic constraint $(k_2/k_1)^\alpha = 0.1364 = 12/88$:

\begin{equation}
\mathcal{J}(\mathbf{p})=
\sum_{i=1}^{N}
\left[
f_A^{\mathrm{exp}}(t_i)-f_A^{\mathrm{model}}(t_i;\mathbf{p})
\right]^2
+
\left[
\omega\left(
\left(\frac{k_2}{k_1}\right)^\alpha-0.1364
\right)
\right]^2,
\end{equation}

\noindent
where $\omega \gg 0$ is a positive penalty parameter (or regularization factor) chosen to ensure that the equilibrium distribution satisfies the empirical thermodynamic constraint. The uncertainties associated with the optimized parameters $\mathbf{p}=[f_A(0),k_1,k_2,\alpha]^T$
were estimated from the covariance matrix:

\begin{equation}
\mathbf{\Sigma} = \sigma^2 \left(\mathbf{J}^{T}\mathbf{J}\right)^{+},
\end{equation}

\noindent
where $(\cdot)^{+}$ denotes the Moore--Penrose pseudoinverse, $\mathbf{J}$ is the Jacobian matrix evaluated at the optimal solution, and $\sigma^2$ represents the residual mean squared error.
The standard errors were obtained as the square roots of the diagonal elements of $\mathbf{\Sigma}$.

Table~\ref{tab:resultados_ajuste_classico} presents the estimated values for $f_A(0)$, $k_1$, $k_2$, and $\alpha$. The estimated value of $\alpha \approx 0.21$ demonstrates a significant deviation from classical exponential behavior ($\alpha=1$), indicating dynamics governed by the memory effects characteristic of the Caputo fractional derivative. The ratio $(k_2/k_1)^\alpha$ remained close to the value specified by the imposed penalty. Additionally, excellent agreement is observed between the slopes obtained in the log--log representation of the experimental data and those of the fitted solution.

The high uncertainties associated with the parameters $k_1$ and $k_2$ reflect a strong correlation between them, indicating that the experimental data lack sufficient information to identify these two parameters completely independently. This limitation is typical of models in which different parameter combinations can yield very similar responses, as the analytical solution depends on $k_1$ and $k_2$ via the combinations $k_1^\alpha+k_2^\alpha$ and $(k_2/k_1)^\alpha$. Consequently, the model is substantially more sensitive to $\alpha$ than to the individual kinetic constants, a disparity in sensitivity that stems from the underlying mathematical structure of the model itself.

\begin{table}[H]
\centering
\caption{Results of the Fractional Model Fit.}
\label{tab:resultados_ajuste_classico}
\begin{tabular}{l|c}
\hline\hline
\textbf{Parameter} & \textbf{Estimated Value} \\ \hline
$k_1$ & $(5 \pm 14)\,\mathrm{ms}^{-1}$ \\
$k_2$ & $(0.54 \pm 1.30) \times 10^{-3}\,\mathrm{ms}^{-1}$ \\ 
$\alpha$ & $0.22 \pm 0.01$ \\ 
$(k_1/k_2)^\alpha$ & $0.137 \pm 0.004$\\
& (Expected: $0.136$) \\
Log--log slope  & $-0.1050\pm 0.0041$\\
& (Expected: $-0.1034$) \\ \hline\hline
\end{tabular}
\end{table}

\begin{figure}[H]
\centering
\includegraphics[width=0.85\textwidth]{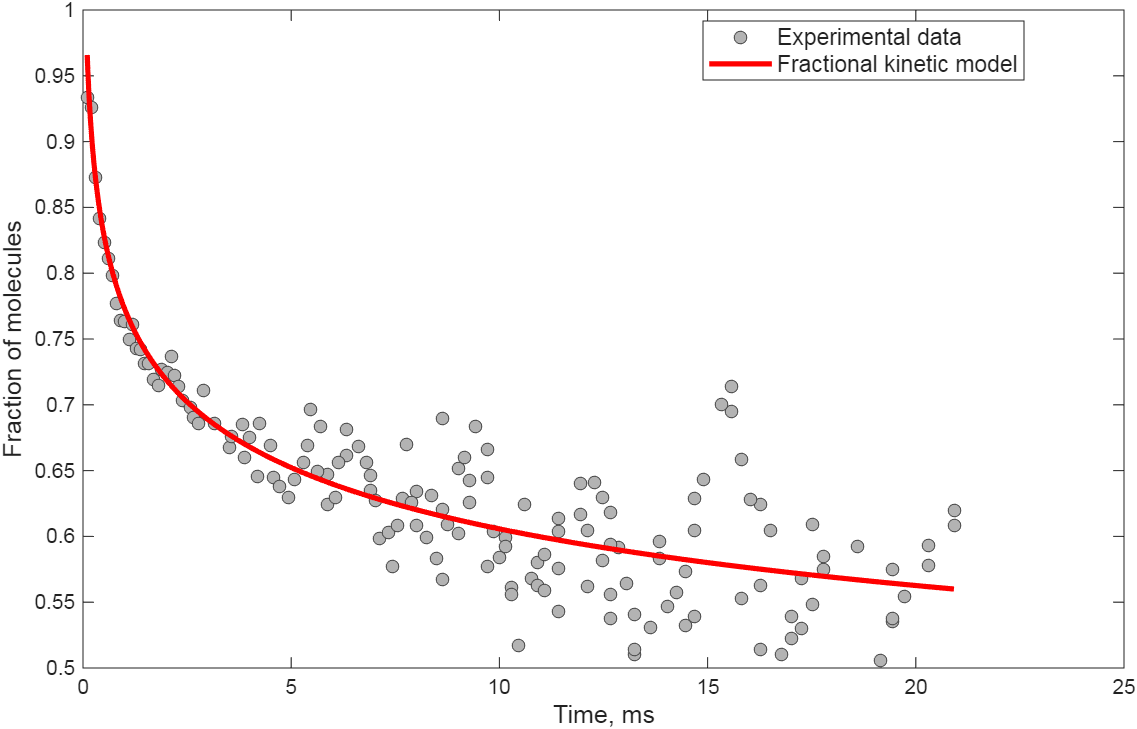}
\caption{Temporal evolution of the population fraction of closed conformers of the TbQSOX enzyme on a standard linear scale. Gray circles represent the digitized experimental data reported by Grossman-Haham et al.~\cite{grossman2018slow}. The solid red line denotes the analytical fit of the unconstrained fractional kinetic model based on the generalized Mittag--Leffler function ($E_{\alpha}$). }
\label{fig:ajuste_linear_individual}
\end{figure}

\begin{figure}[H]
\centering
\includegraphics[width=0.85\textwidth]{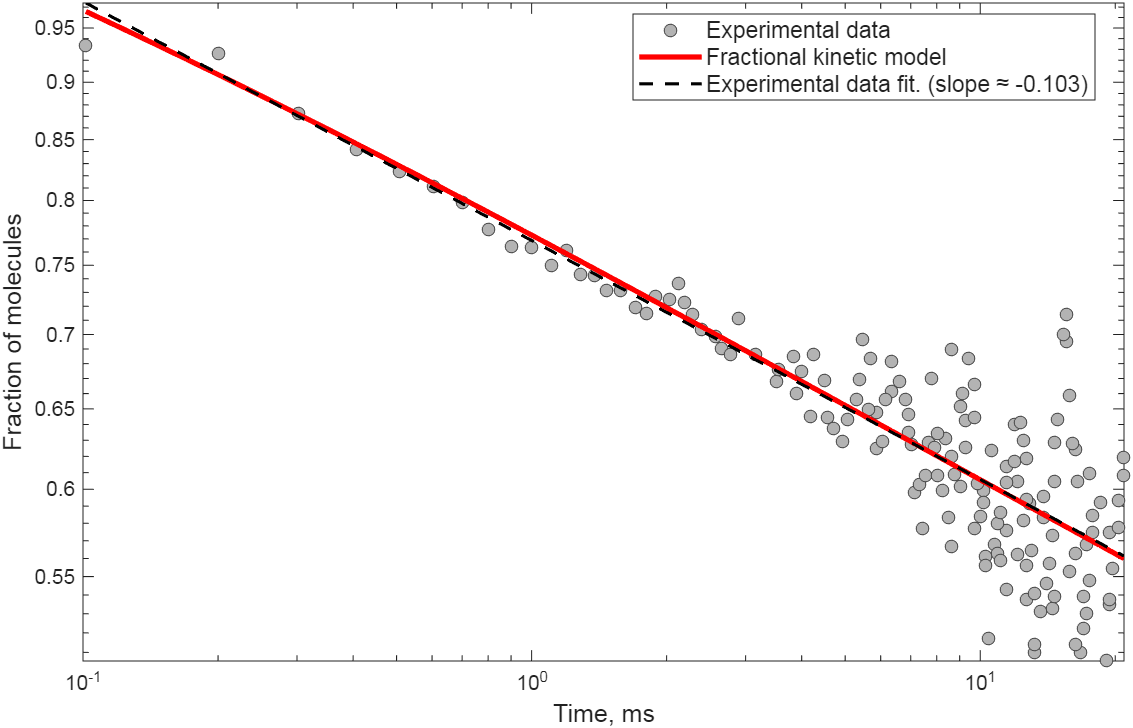}
\caption{Log--log kinetic diagnostic analysis of the closed conformer fraction. Gray circles represent the data extracted from Grossman-Haham et al.~\cite{grossman2018slow}. The solid red line illustrates the asymptotic behavior predicted by the optimized fractional model, while the dashed black line denotes the empirical trend line of the data.}
\label{fig:ajuste_loglog_individual}
\end{figure}

Figure~\ref{fig:ajuste_linear_individual} illustrates the temporal evolution of the population fraction of closed-state conformers of the TbQSOX enzyme. Figure~\ref{fig:ajuste_loglog_individual} presents the same experimental data and the analytical solution of the fractional model, plotted on a log--log scale ($\log_{10} t$ versus $\log_{10}f_A$), which serves as a definitive test for identifying power-law dynamics. This figure clearly demonstrates the capability of the proposed model to reproduce the non-exponential relaxation profile of the system.

\begin{figure}[H]
\centering
\includegraphics[width=0.85\textwidth]{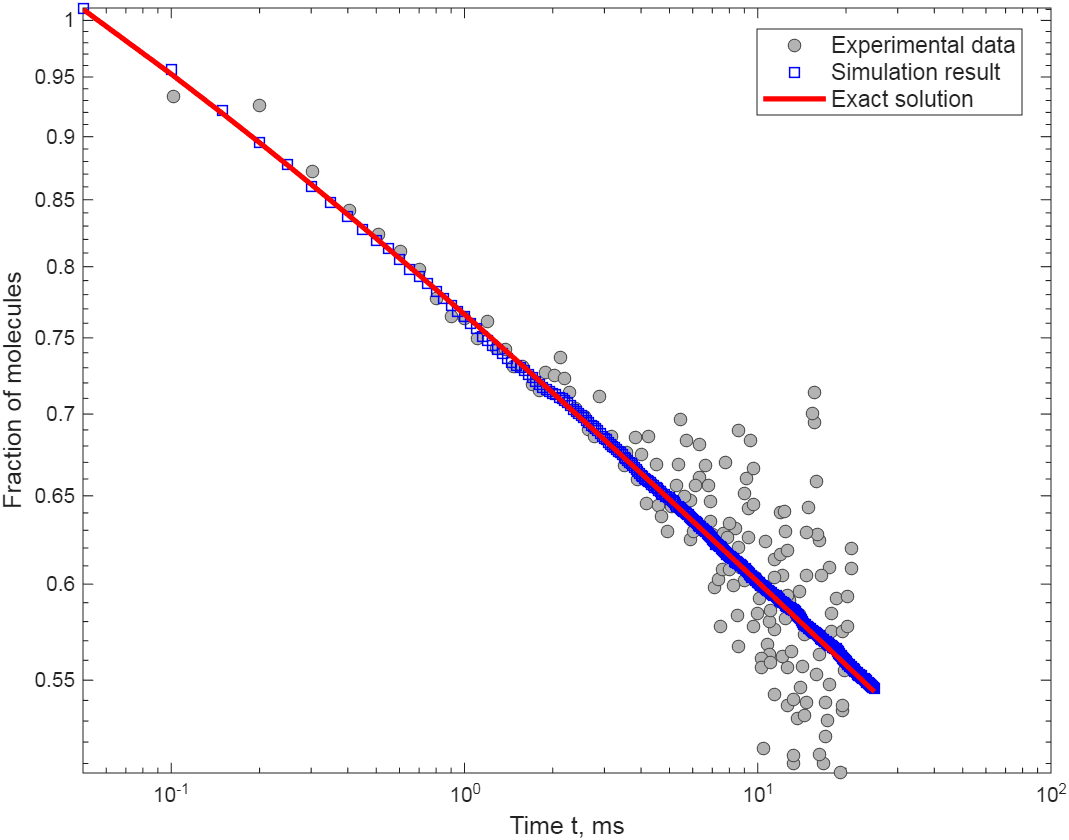}
\caption{Log--log plot comparing the experimental kinetic data of the TbQSOX enzyme (gray filled circles) with the fractional kinetic model ($\alpha = 0.2$). The solid red line represents the exact analytical solution governed by the Mittag--Leffler function, while the open blue squares depict the results of the exact time-driven fractional Monte Carlo simulation.}
\label{fig:trajetoria_unica}
\end{figure}

Figure~\ref{fig:trajetoria_unica} presents the results of a time-driven Monte Carlo simulation of the fractional kinetic process $A \rightleftharpoons B$ for a single stochastic trajectory, utilizing the kinetic parameters detailed in Table~\ref{tab:resultados_ajuste_classico}. These results are compared with the exact analytical solution governed by the Mittag--Leffler function ($\alpha = 0.22$). As expected for a single-trajectory realization, the stochastic process exhibits highly noisy behavior. Nevertheless, clear qualitative agreement is observed between the simulated Monte Carlo trajectory and the analytical fractional solution. As depicted on a log--log scale, the overarching trend of the stochastic simulation (blue squares) closely matches the exact analytical curve (red line) across all timescales. This strong alignment validates the numerical stability of our algorithm (equation (\ref{eq:Pexact_global2})), demonstrating that it successfully circumvents the theoretical hazard rate divergence as $t \to 0$ (equation (\ref{complica})).

A systematic study of the proposed method for the case of an irreversible process is provided in Reference~\cite{TCAM}. That work examined numerical convergence with respect to the time step ($\Delta t \to 0$), structural convergence in the limit $\alpha \to 1$, and statistical convergence with respect to the ensemble size ($M \to \infty$). The latter analysis is reproduced here for the irreversible process. Convergence in the limit $\alpha \to 1$ is also confirmed here for the same reasons detailed in Reference~\cite{TCAM}.

\begin{figure}[htbp]
\centering
\includegraphics[width=0.85\textwidth]{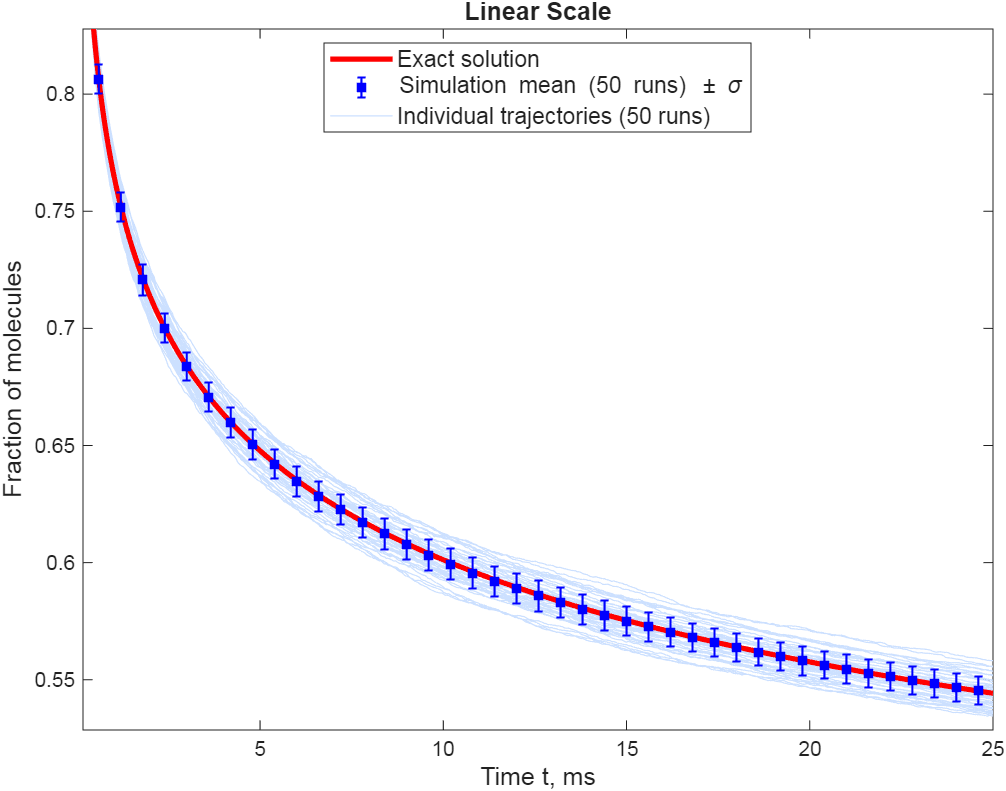}
\caption{Statistical analysis of an ensemble of $M = 50$ independent Monte Carlo trajectories (thin blue lines) for the fractional kinetic system ($\alpha = 0.2$). The ensemble mean of the simulations (blue squares) is compared with the exact analytical solution (solid red line). The vertical bars indicate the sample standard deviation ($\pm 1\sigma$) calculated over the temporal evolution, demonstrating the convergence of the stochastic estimator toward the macroscopic solution.}
\label{fig:ensemble_erros}
\end{figure}

The fluctuations observed in the Monte Carlo simulation points are statistically expected, given the intrinsically noisy nature of the stochastic process and the fact that only a single trajectory is considered. Figure~\ref{fig:ensemble_erros} presents the ensemble average over $M=50$ independent trajectories of the fractional unimolecular decay process, obtained via time-driven Monte Carlo simulation, alongside error bars corresponding to the standard deviation $\sigma_N(t)$.

In the Monte Carlo formulation of stochastic chemical kinetics, the physically relevant observable is the ensemble-averaged trajectory rather than an individual realization. Let $N^{(m)}(t)$ denote the $m$-th independent stochastic realization of the process. The empirical ensemble mean is defined by the estimator:

\begin{equation}
\langle N(t) \rangle_M = \frac{1}{M} \sum_{m=1}^{M} N^{(m)}(t).
\end{equation}

For a fixed time $t$, the set $\{N^{(m)}(t)\}_{m=1}^{M}$ consists of independent and identically distributed (i.i.d.) random variables with finite variance $\sigma^2(t)$. By the strong law of large numbers, the empirical mean converges almost surely to the theoretical expectation:

\begin{equation}
\langle N(t) \rangle_M \longrightarrow \mathbb{E}[N(t)] \quad \text{when\hspace{0.2cm}} M \to \infty,
\end{equation}

\noindent
which guarantees the statistical consistency of the Monte Carlo estimator.

\section{Conclusion}

This work presented a unified analytical and stochastic framework based on fractional calculus to model and interpret the anomalous kinetic behavior observed in the conformational transitions of the TbQSOX enzyme.

Within the mechanistic interpretation proposed by Grossman--Haham \emph{et al.}~\cite{grossman2018slow}, the anomalous kinetics arises from a rugged free-energy landscape, where the molecule experiences repeated transient trapping in numerous local minima before reaching the opposite conformational state. Consequently, the conformational landscape is explored inefficiently, giving rise to heavy-tailed waiting-time distributions and anomalously slow relaxation dynamics.

To model this observation theoretically, the authors in ref. ~\cite{grossman2018slow} employed a subdiffusive model governed by a diffusion equation whose coefficient, $D(t)$, undergoes temporal attenuation according to the relation $D(t) \sim t^{\alpha-1}$. Within the CTRW formalism used here this physical picture is naturally captured by the fractional exponent $\alpha$, which quantifies the degree of temporal memory associated with the trapping process.

In contrast to the simple exponential decay predicted by traditional models, the linear behavior observed in log--log space constitutes mathematical evidence that the system obeys a power-law relation.

The fractional model postulates that the waiting times between reaction events follow $\psi(t) \sim t^{-(1+\alpha)}$. Within this framework, the physical argument proposed by Grossman--Haham \emph{et al.} is naturally accommodated; this anomalous behavior emerges from dynamic disorder intrinsic to the enzyme's multidomain architecture, where the molecule becomes transiently trapped in multiple local minima while attempting to reconfigure. In the continuous-time random walk (CTRW) formalism, this physical trapping phenomenon is mathematically captured by heavy-tailed (power-law) waiting-time distributions.

The quantitative application of this model to the experimental data for the TbQSOX enzyme demonstrated excellent agreement, accurately capturing both the non-exponential relaxation profile and the dynamic self-similarity evidenced on a log--log scale. The estimated value of the fractional exponent ($\alpha \approx 0.22$) provided an unequivocal quantitative signature of subdiffusion and a marked departure from classical exponential kinetics ($\alpha = 1$).

Consequently, the proposed fractional stochastic model not only provides a robust fit to the data but also establishes a rigorous mathematical foundation for the underlying biological hypothesis, demonstrating its superiority over previously employed models.

\bibliographystyle{ieeetr}
\bibliography{artbiblio,bibforaeq}

\end{document}